# Schwarzschild Black Hole Lives to Fight Another Day – Comment on the paper J. Math. Phys. 50, 042502 (2009) by A. Mitra

Prasun K. Kundu

*Department of Mathematics and Statistics, University of Maryland, Baltimore County, Baltimore, MD, USA*

Abstract

In a comment published several years ago in this Journal [J. Math. Phys. **50**, 042502 (2009)] Mitra has claimed to prove that a neutral point particle in general relativity as described by the Schwarzschild metric must have zero gravitational mass, i.e. the mass parameter $M_0$ of a Schwarzschild black hole necessarily vanishes. It is shown that the purported proof is incorrect. The error stems from a basic misunderstanding of the mathematical description of coordinate volume element in a differentiable manifold.

Several years ago in a paper published in this Journal by Mitra [1] as a comment on an earlier paper by Castro [2] it was claimed that in general relativity a neutral point particle must have zero gravitational mass. The claim is indeed an extraordinary one, since it would imply the non-existence of a Schwarzschild black hole as a physically realizable object in nature. This would, in effect, upend much of the conventional wisdom regarding the physics of relativistic astrophysical objects, such as quasars and active galactic nuclei. It has long been known [3,4] that the Schwarzschild space-time with the metric (in units $G = c = 1$)

$$ds^2 = (1 - 2M_0/R)dT^2 - (1 - 2M_0/R)^{-1}dR^2 - R^2(d\theta^2 + \sin^2\theta d\phi^2)$$

represents the gravitational field of a neutral point particle of gravitational mass $M_0$ located at the curvature singularity $R = 0$. The apparent singularity at $R = 2M_0$ is merely an artifact of the standard Schwarzschild coordinates $(T, R, \theta, \phi)$ (which Mitra refers to as Hilbert coordinates), that can be removed by going over to the Kruskal extension. It is in fact a nonsingular 3-d subspace of the Schwarzschild space-time (of zero proper volume, whose $T$ = const 2-d slices have area $16\pi M_0^2$) representing an event horizon for distant stationary observers [3]. When $M_0 = 0$, it simply reduces to the empty flat Minkowski space-time. Without entering into a detailed critique of the exchange between the two authors, we here show that the proof presented by Mitra in [1] is erroneous.

In Sec. II of the paper [1] Mitra proceeds to employ the familiar formula for the invariant volume element in a (pseudo-)Riemannian 4-manifold with metric tensor $g_{\alpha\beta}$ in the context of the geometry of Schwarzschild space-time. Specifically, he considers the transformation of the invariant 4-volume element in Schwarzschild geometry between the Schwarzschild coordinates $(T, R, \theta, \phi)$ and the Eddington-Finkelstein coordinates $(T_*, R, \theta, \phi)$. While his Eq. (8) stating this invariance $\sqrt{-g'}d^4x' = \sqrt{-g}d^4x$ is essentially correct, the inference drawn from it, namely the relation $dT_* = dT$ *is not*. On the other hand from Eq. (5) defining $T_*$ (which is correct modulo misplaced parentheses), it follows that $dT_* = dT \pm [2M_0/(R - 2M_0)]dR$. The attempt to reconcile these two relations is what leads to the absurd conclusion that the mass parameter $M_0$ appearing in the metric should vanish. The error appears to stem from a simple misunderstanding of the mathematical meaning of volume element in differential geometry and its transformation under a diffeomorphism (or equivalently coordinate transformation). In order to present the essence of the argument we analyze the problem below in two dimensions analogous to the 2-d subspace spanned by the coordinates $T, R$ leaving aside the angular coordinates.

Consider a transformation between general curvilinear coordinates $(x^1, x^2)$ and $(x'^1, x'^2)$ in a 2-d Riemannian manifold. In the language of differential geometry [3,4] the coordinate differentials $dx^1$ and $dx^2$ in a differentiable manifold are not scalar quantities but 1-forms and the area element is a 2-form $d^2x = dx^1 \wedge dx^2$, where $\wedge$ denotes the exterior product ('wedge product'). Under a coordinate transformation $x'^\alpha = x'^\alpha(x^\beta)$ the coordinate area element transforms as

$$d^2x' = dx'^1 \wedge dx'^2 = J\, dx^1 \wedge dx^2 = J\, d^2x$$

where $J = \det|\partial x'/\partial x|$ is the Jacobian of the transformation. From the tensor transformation law of the metric tensor $g_{\alpha\beta}$ it follows that the determinant of the metric transforms as $g' = J^{-2}g$, in turn implying the invariant nature of the proper area element, i.e. $\sqrt{g'}d^2x' = \sqrt{g}d^2x$. It is a mathematical identity valid for all metrics and it should be clear that, by itself, it cannot yield any further restriction on $g_{\alpha\beta}$ as claimed in Mitra's paper [1].

The error in Mitra's reasoning arises in his treating the wedge product implied in the definition of volume as a simple algebraic product of the coordinate differentials. For the particular class of transformation $x'^1 = x^1 + f(x^2)$, $x'^2 = x^2$ which includes the one considered by Mitra [Eq. (5)], $J = 1$ and indeed $dx'^1 \wedge dx'^2 = [dx^1 + (df/dx^2)dx^2] \wedge dx^2 = dx^1 \wedge dx^2$ for any choice of $f$, just as one would expect, since $dx^2 \wedge dx^2$ identically vanishes. It is obviously incorrect to conclude from this relation that $dx'^1 = dx^1$ when $dx'^2 = dx^2$ as Mitra has done. Omitting the wedge product is a very common abuse of notation in the general relativity literature. However, it should be taken into account carefully in mathematical manipulations as in Mitra's Eq. (8) and ignoring it could easily yield nonsensical result.

In summary, we have shown that the proof of the vanishing of the mass parameter $M_0$ of the Schwarzschild metric given by Mitra [1] is incorrect. This invalidates his central claim that a Schwarzschild black hole of nonzero mass does not exist in nature and is merely a mathematical curiosity. It also calls into question his various other more physically based arguments that apparently reached the same conclusion [5] and casts serious doubt on his proposal of a black hole alternative ('Eternally Collapsing Objects'). Unfortunately, this dubious claim has further propagated in the literature (e.g. in [6]). Indeed the Schwarzschild space-time is a valid solution of the vacuum Einstein equations for *all* values of $M_0$ (including negative ones!). Presumably, only the positive $M_0$ solutions representing black holes are physically realized through gravitational collapse of 'normal' matter satisfying suitable positive energy conditions.